%% file: main.tex
\documentclass[10pt,conference]{IEEEtran}
\usepackage{cite}
\usepackage{listings}
\usepackage{subcaption}
\input{format}

\newcommand{\lof}[1][]{%
\ifthenelse{\equal{#1}{}}{\emph{LoF}\xspace}{\emph{LoF-{#1}}\xspace}%
}

\newcommand{\loffull}{Level of Fidelity (LoF)}

\newcommand{\lofs}{\emph{LoFs}\xspace}

\begin{document}

\title{Coupled Requirements-driven Testing of CPS in Simulation and Reality}
\title{Coupled Requirements-driven Testing of CPS:\\ From Simulation To Reality}

\author{\IEEEauthorblockN{Ankit Agrawal}
\IEEEauthorblockA{\textit{Department of Computer Science} \\
\textit{St. Louis University, USA}\\
ankit.agrawal.1@slu.edu}
\and
\IEEEauthorblockN{Philipp Zech}
\IEEEauthorblockA{\textit{Department of Computer Science} \\
\textit{University of Innsbruck, Austria}\\
philipp.zech@uibk.ac.at}
\and
\IEEEauthorblockN{Michael Vierhauser}
\IEEEauthorblockA{\textit{Department of Computer Science} \\
\textit{University of Innsbruck, Austria} \\
michael.vierhauser@uibk.ac.at}
\and

}


\maketitle

\input{sec_00_abstract}

\vspace{1em}
\begin{IEEEkeywords}
Testing, Simulation, Requirements, Small Unmanned Aerial Systems, Process, Cyber-Physical Systems
\end{IEEEkeywords}


\input{sec_01_introduction}
\input{sec_02_motivation}

\input{sec_03_process}

\input{sec_04_instance}
\input{sec_05_roadmap}
\input{sec_06_conclusion}




\balance


\bibliographystyle{IEEEtran}
\bibliography{refs.bib}

\end{document}

%% file: format.tex
\usepackage[T1]{fontenc}
\usepackage{booktabs} 
\usepackage{xcolor}
\usepackage{url}
\usepackage{multirow}
\usepackage{rotating}
\usepackage{array}
\usepackage{color}
\usepackage{wasysym}
\usepackage{verbatim}
\usepackage{bigstrut}
\usepackage{multirow}
\usepackage{setspace}
\usepackage{tabularx}
\usepackage{url}
\usepackage{xspace}
\usepackage{pifont}
\usepackage{subcaption}
\usepackage[font=footnotesize]{caption}
\usepackage{array,booktabs}
\usepackage{soul}
\usepackage{vcell}
\usepackage{pifont}
\usepackage{afterpage}
\usepackage{rotating}
\usepackage{wrapfig}
\usepackage{caption}
\usepackage{graphicx}
\usepackage{caption}
\usepackage{enumitem}
\usepackage{lipsum}
\usepackage{balance}
\usepackage[ruled,vlined]{algorithm2e}
\usepackage{algpseudocode}
\usepackage{tikz}
\usepackage{hyperref}

\usepackage[colorinlistoftodos,prependcaption,textsize=tiny]{todonotes}

\newcolumntype{L}[1]{>{\raggedright\let\newline\\\arraybackslash}p{#1}} 
\newcolumntype{C}[1]{>{\centering\let\newline\\\arraybackslash}p{#1}} 
\newcolumntype{M}[1]{>{\centering\arraybackslash}m{#1}}
\newcolumntype{R}[1]{>{\raggedleft\let\newline\\\arraybackslash}p{#1}} 

\newcommand{\citesec}[1]{Section~\ref{sec:#1}}
\newcommand{\citefig}[1]{Fig.~\ref{fig:#1}}

\newcommand{\citetable}[1]{{Table}~\ref{tab:#1}}

\newcommand{\etal}[1][]{%
\ifthenelse{\equal{#1}{}}{\emph{et~al.}\xspace}{\emph{et~al.}~\cite{#1}\xspace}%
}

\definecolor{alizarin}{rgb}{0.82, 0.1, 0.26}

%% file: sec_00_abstract.tex
\begin{abstract}
Failures in safety-critical Cyber-Physical Systems (CPS), both software and hardware-related, can lead to severe incidents impacting physical infrastructure or even harming humans. As a result, extensive simulations and field tests need to be conducted, as part of the verification and validation of system requirements, to ensure system safety. However, current simulation and field testing practices, particularly in the domain of small Unmanned Aerial Systems (sUAS), are ad-hoc and lack a thorough, structured testing process. Furthermore,  there is a dearth of standard processes and methodologies to inform the design of comprehensive simulation and field tests. This gap in the testing process leads to the deployment of sUAS applications that are: (a) tested in simulation environments which do not adequately capture the real-world complexity, such as environmental factors, due to a lack of tool support; (b) not subjected to a comprehensive range of scenarios during simulation testing to validate the system requirements, due to the absence of a process defining the relationship between requirements and simulation tests; and (c) not analyzed through standard safety analysis processes, because of missing traceability between simulation testing artifacts and safety analysis artifacts. To address these issues, we have developed an initial framework for validating CPS, specifically focusing on sUAS and robotic applications. We demonstrate the suitability of our framework by applying it to an example from the sUAS domain. Our preliminary results confirm the applicability of our framework. We conclude with a research roadmap to outline our next research goals along with our current proposal.
\end{abstract}

%% file: sec_01_introduction.tex
\section{Introduction}
\label{sec:intro}

Software Testing is a well-established research area, with mature processes, tools, and techniques~\cite{orso2014software}. Over the years, a multitude of testing methods, such as model-based testing~\cite{utting2010practical}, mutation testing~\cite{jia2010analysis}, metamorphic testing~\cite{chen2018metamorphic,chen2020metamorphic
}, and search-based testing~\cite{harman2009theoretical,afzal2009systematic
}, to name but a few, have emerged. 
Furthermore, the ISO 29119 standard~\cite{patricio2021study}, for example, covers testing processes, documentation, techniques, and guidelines for thorough software application testing.
Depending on the application domain and context, software testing contributes to detecting bugs and errors in the System under Test (SuT), uncovering performance issues, and identifying missing or incorrectly implemented requirements contributing to assessing the quality of a software product. In case of safety-critical applications, testing and test results are leveraged as evidence for safety assurance, arguing about system safety, for example, as part of Safety Assurance Cases (SACs)~\cite{hawkins2013assurance,hecht1997quality}.

Particularly for Cyber-Physical Systems (CPS), such as robotic applications, operating as part of smart production environments, or small Uncrewed Aerial Systems (sUAS), used in search-and-rescue operations, failures during operation can result in severe incidents harming physical infrastructure, the surrounding environment, or even humans.
It is, therefore, crucial to specify safety requirements, implement mitigations, and moreover, establish a thorough testing and validation process~\cite{ali2015u}.
In CPS, this is commonly achieved with a combination of traditional testing activities, simulations conducted in a high-fidelity simulator, as well as Hardware in the Loop (HITL), and field tests.
While, for example, for the automotive domain, thorough test and safety standards have been established~\cite{bringmann2008model}, other areas still lack a thorough testing and validation process.
This is particularly true in the domain of sUAS and small unmanned ground vehicles (UGV). While a multitude of different simulation environments exist, providing support for testing individual aspects of a CPS, such as collision avoidance, path planning, and vision-based navigation,   an overarching and holistic testing and validation process is still missing.
Particularly due to the fact that thorough validation and/or certification is not mandated by standards or regulations, these types of systems often only employ ad-hoc testing strategies.
Furthermore, in this context, environmental conditions, play a particular role, as these systems need to operate in real-world scenarios, with diverse combinations, of, for example, wind conditions, weather, lighting, terrain geographical information.
Simulators, however, are typically limited in what can be simulated and tested, and the level of fidelity, of how well a simulation represents the real world and the behavior of the hardware, greatly varies depending on the simulation environment~\cite{simgap2}.
This results in situations where a system that has been thoroughly tested, with numerous simulation runs being executed, still deviates from its specified behavior when used in the ``real world'' due to insufficient or low-fidelity simulations being executed~\cite{agrawal2023requirements}.

To this end, we have created an initial framework for testing  CPS, with a particular emphasis on sUAS and robotic applications that strongly rely on simulations, HITL, and ultimately real-world field testing. Based on our research in the domain of sUAS applications, and the lack of stringent regulations in this area (e.g., compared to the automotive domain~\cite{bringmann2008model}), we found that there is a dearth of structured testing concepts that incorporate requirements, diverse validation properties, and multi-facet integration testing (e.g., different simulations, in conjunction with field tests within a structured well-defined test context).
Based on these initial insights, and our own experiences, where we have faced similar issues, we have derived an initial testing process and framework.

The contributions of this RE@Next paper are as follows: First, we discuss existing work and provide a motivating example for the need for additional research. Second, we lay out a structured testing process, and the artifacts/steps necessary in the different phases.
Third, we provide an application example, of how the process would be executed for performing tests for a sUAS system. Fourth, we present a research roadmap of open questions and avenues that require additional research.

 The remainder of this paper is structured as follows. In~\citesec{background}, we provide a motivating example and discuss related work, and identify gaps. In Sections~\ref{sec:solution} and~\ref{sec:instance} we provide an overview of our initial testing framework and process, and how it can be applied. Finally, we present our research roadmap in \citesec{roadmap}, before we conclude in \citesec{conclusion}.




%% file: sec_02_motivation.tex
\section{Background \& Motivating Example }
\label{sec:background}
In the following, we provide a brief introduction to the area of simulator-based testing for sUAS and present a motivating example for establishing a thorough workflow and process.

\noindent$\bullet$ \textbf{sUAS Simulation Tools and Capabilities:}
CPS in general, and sUAS in particular, combine hardware, software, and human interaction components, under complex and uncertain environments. Wind, for example, is a significant environmental factor impacting sUAS operations, leading to a range of hazards and unpredictable behaviors \cite{vierhauser2021hazard}. For instance, Meta's Aquila project, an initiative to develop a fixed-wing UAS capable of distributing internet, experienced a setback when the UAS could not adequately handle an unexpected wind gust, resulting in a severe crash \cite{Facebook45:online}.

Despite numerous specialized and open-source sUAS simulation tools, realistically testing sUAS behavior in simulated windy conditions is challenging. The nature of wind is highly dynamic, as its velocity and direction are influenced by the complex geometry of the environment. However, widely popular simulation tools, including Gazebo\cite{gazebo} and AirSim \cite{shah2018airsim}, only allow for the simulation of basic wind conditions. These include constant wind in a specific direction applying uniform wind pressure throughout the flight, or turbulent wind based on a statistical model without considering the complex geometry of the environment. These simplifications in the wind simulations create a significant gap between the testing results in simulations and the actual behavior of sUAS during live operations. Further, simulation tools overlook other crucial factors such as GPS signal fluctuations, communication quality degradation, and radio interference, which significantly affect the real-world sUAS operations.


\noindent$\bullet$ \textbf{CPS-Testing:} In the context of CPS, a wide variety of testing and validation approaches exist to validate the behavior of sUAS before field deployments. Techniques range from assessing the accuracy of onboard AI models, to applying software testing techniques such as Fuzzy and Metamorphic testing \cite{chen2018metamorphic}, to automatically generate new diverse test cases. However, most of these approaches have only focused on validating certain aspects of the SuT, often employed in isolation, focusing on individual subsystems or components, rather than the entire deployable sUAS. For instance, consider an autonomous sUAS designed for wildfire monitoring and surveillance. In such scenarios, a sUAS's obstacle avoidance system must swiftly navigate around unpredictable obstacles in the air, while its imaging system must keep a steady focus on hotspots, despite the sUAS' movements. Besides individual functionality, all components of the system need to operate flawlessly in unison, which is often left untested by standard testing procedures.

In a recent survey, Abbaspour~\etal[abbaspour2015survey]
created a categorization and comparison of CPS testing levels and analyzed them in the context of the V-Model. Two of the main challenges that were uncovered are (1)  ``Extra Functional Properties'' that need to be taken into account, and moreover (2) providing support for (automated) evaluation of the test results.
The latter, however, is an integral aspect of a holistic testing process, particularly when simulation is involved, as typically a large number of simulation runs are executed, with hundreds or even thousands of configuration parameters involved. For instance, during simulation testing, it is not always immediately clear if, and to what extent, a simulation has in fact executed correctly, or if any (minor) deviations from the expected behavior occurred. Similar challenges for CPS testing were reported by Zhou~\etal[zhou2018review], further emphasizing uncertainties and the interactions with  physical systems, and the need to `` simulate and co-simulate many parallel physical processes to analyze
the continuous interaction of the user, controller, and physical environment and to automatically identify the worst-case scenarios''.

\noindent$\bullet$ \textbf{Model-based Testing:} Model-based testing (MBT) is a software testing approach that treats models as first-class citizens~\cite{pretschner2005model}. 
It utilizes models to depict the anticipated behavior of a SuT, e.g., a sUAS. Significantly, rather than depending on manual test case development, MBT utilizes a model that delineates the characteristics and functionality of the SuT, with the objective of automatically generating test 
cases from this model. MBT enables the efficient exploration of the search space to identify important scenarios for testing purposes~\cite{dalal1999model}. 
MTB defines a slim and efficient process consisting of three core activities:
First, during \emph{Model Creation}, testers construct a model that accurately portrays the intended functionality of the SuT. The model may take the form of a graphical representation, a mathematical formulation, or a hybrid of the two. Second, during \emph{Test Case Generation}, test cases are inferred automatically from the model. Specialized tools can be utilized to evaluate the model and generate test scenarios; however, tests may also be modeled consistently in accordance with the models of the system. Finally, during \emph{Execution and Evaluation}, test cases, whether manually created or automatically inferred, are executed against the implemented SuT. Test results are then compared to the predefined oracles (i.e., expected test outcome). All inconsistencies or malfunctions are documented for subsequent examination and resolution.

The use of MBT offers several advantages in terms of efficiency, validity, relevance of test cases, and maintainability~\cite{utting2010practical}. MBT allows for efficient test design, ensuring comprehensive 
coverage and consistency in accordance with the SuT. Additionally, MBT facilitates easy adaptation of test cases by reflecting changes in models, resulting in quick 
updates and maintainability. Nevertheless, MBT presents several problems, including the necessity for meticulous model creation and the reliance on specialist tools. 
Moreover, it may not be appropriate for all categories of projects or testing situations. Yet, over the last decades, MBT has been successfully applied for safety-critical testing of software systems, with a strong focus on the automotive and railway domains. However, its application for safety-critical testing in the avionics domain remains substantially unexplored so far~\cite{gurbuz2018model}.

\bigskip
The goal of our work is to leverage these existing techniques to advance and streamline the testing process.
Drawing from the challenges in existing CPS testing, and our experience with CPS and sUAS simulations, we have developed a preliminary framework. The next section outlines this framework, detailing the process, model, and potential applications.

%% file: sec_03_process.tex
\section{Towards a structured and flexible CPS Simulation Testing Process}
\label{sec:solution}

Commensurate with our discussion from~\citesec{background}, we identified a significant gap concerning structured testing and requirements validation support, in combination with safety assurance tasks.

To address these shortcomings we have created an initial version of a systematic testing and validation process for simulation which is outlined in~\citefig{sim_process}. 
The process is guided by the ISO 29119 standard on Software Testing, with a particular focus on (1) \emph{how simulator capabilities and simulation needs} are incorporated in the testing process; (2) how traceability can be established between requirements to be validated and test results; and (3) how different levels of fidelity can be reflected in the testing process.
As part of this, we have further identified several key artifacts and models. In the following, we provide a brief overview of these artifacts, their purpose and contents (cf.~\citetable{overview}), and how they are used in the overall testing and validation process. In the subsequent discussion and research roadmap (cf.~\citesec{roadmap}), we further lay out particular aspects that require further attention and investigation.

\subsection{Requirements Modelling}
As a precursor to establishing a successful and thorough testing and validation process, system requirements, both functional and non-functional need to be clearly defined and documented. 
\emph{Modeling requirements} is a well-established practice in the field of requirements engineering for both ``traditional'' software systems, as well as safety-critical and cyber-physical systems. 
Approaches range from dedicated requirements modeling languages and frameworks, such as i*~\cite{dalpiaz2016istar}, Tropos~\cite{giorgini2004tropos}, and KAOS~\cite{dardenne1993goal}, to formal approaches, such as temporal logic~\cite{greenspan1994formal}, 
or the Z notation~\cite{awan2020formal}.
In our framework, we do not prescribe a specific language, notation, or technique, as long as the used approach facilitates the structured capturing and description of relevant requirements.\newline

Additionally, pertaining to CPS in general, and sUAS applications in particular, their operation in close proximity to humans, in real-world environments, requires several safety-related aspects, and hence safety requirements, to be taken into consideration.
Therefore, to facilitate proper testing and validation, as a precursor, we assume the existence of a \emph{Requirements Model} that documents these in a structured manner, and a \emph{Verification \& Validation Model (V\&V Model)} that contains information on properties to be validated.  In particular, these properties should include both (a) test properties such as maintaining minimum sUAS battery during the mission, keeping flight path deviation percentage under a threshold level, and not breaching a restricted airspace, and (b) environment properties such as windy conditions, lighting conditions, and GPS satellite availability in the environment under which the SuT should maintain those test properties. 
We do not prescribe a specific modeling technique or language, but any technique that facilitates property specification and validation criteria can be used. Examples include formal approaches such as temporal logic
 , UML-based notations~\cite{cimatti2009informal}, or state machines in combination with model-checking~\cite{von1997formal}.

\begin{figure}[t!]
    \centering
    \includegraphics[width=.97\columnwidth]{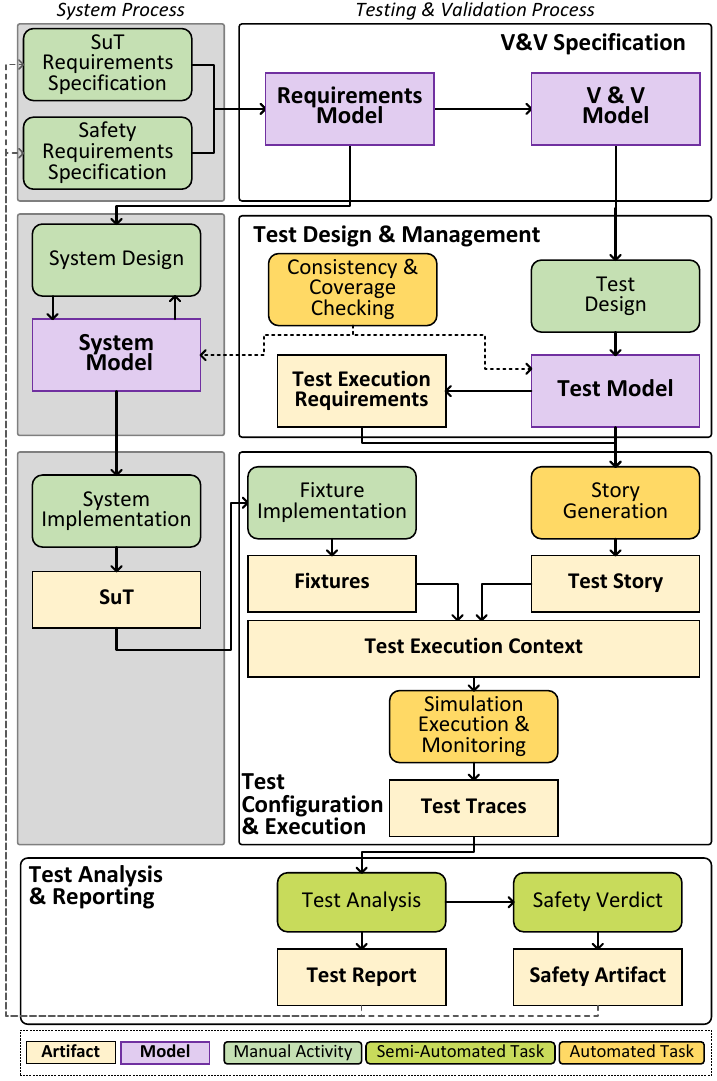}
    \caption{Overview of our process with the constituent artifacts and activities.}
    \label{fig:sim_process}
    \vspace{-1.85em}
\end{figure}

\subsection{Test Design \& Management}
Similar to ISO 29119, the second step is concerned with \emph{Test Design and Management}, i.e., how to develop test cases.
The corresponding \emph{Test Model} contains a description of the scenarios that must be tested and the corresponding configuration needs for the simulation environment.
This initial test specification and design step is largely independent of the concrete test execution, and for example, simulator implementations, only defining a set of capabilities/features that need to be fulfilled by the actual simulation execution environment.
Additionally, during the test design step, each test is associated with a \loffull. The \lof prescribes the level of detail and realism that is required to fully validate the test case, impacting where (i.e., in which test execution context a test has to be executed). In this initial version of the process, we incorporate four fidelity levels, from \lof-0 to \lof-3.

\begin{itemize}%
[leftmargin=1.15em]
    \item [$\bullet$] \emph{{\it \lof-0}:} is the lowest level. The approach has only been validated via component-based testing (e.g., unit tests) and no simulations or external validation has been conducted.

    \item [$\bullet$] \emph{{\it \lof-1}:} describes tests that were executed in a simulation environment. This level is further refined by the respective test execution requirements  (e.g., a simulator needs to fulfill) to successfully execute certain tests.

      \item [$\bullet$] \emph{{\it \lof-2}:}  describes tests that were executed in a HITL setting where certain parts, for example, the drone itself, or parts of it (for example the flight controller) were connected to a simulation environment, and tests were executed collecting runtime data from both (software) simulation environments and hardware parts.

       \item [$\bullet$] \emph{{\it \lof-3}:} finally describes the highest level of fidelity, where test scenarios have been executed in the real world, with physical hardware and human interaction.

\end{itemize}



\subsection{Test Story Generation \& Execution}
Once tests in the Test Model, and the environment requirements are specified, a set of \emph{Test Stories} can be generated. In this part of the process, the abstract and generic test specification (in the test model) transitions to a concrete (and executable) test scenario for a particular execution context, for example, a specific simulation environment.
Depending on the execution context, this requires different types of scenarios to be generated.
In the case of \lof[1] and \lof[2] scripts for configuring and setting up a simulation environment may be created automatically, streamlining the testing process. 
For \lof[3], field test scripts, checklists, and instructions need to be derived. 
It is important to note that tests should always be executed sequentially, from the lowest to the highest \lof.
Particularly for tests that contain safety-critical aspects, a test failing at a lower level also bears a lower safety risk.

\input{tab_overview}

Concrete test stories for the sUAS requirement, for example, specify the characteristics of the 3D environment, such as distinguishing between urban and open settings, and defining the direction, type, and velocity of wind within that environment. Further, these scenarios detail the specific tasks or missions assigned to the sUAS, as well as the initial conditions including the home location of each sUAS in the environment. 

Additionally, to enable testing a SuT in a specific context, \emph{Test Fixtures} need to be created to ensure consistent and controlled test execution.
Test fixtures are commonly used to set up system states and input data needed for test execution, allowing for tests to be repeatable, which is one of the key capabilities of an effective process~\cite{greiler2013automated}. Depending on the actual execution environment, different fixtures need to be created. For example, while using the AirSim simulator, a test fixture could include (a) environment specifications for executing tests, the type of 3D environment — a digital twin model of a city or environment enforcing several navigation constraints; and (b) sUAS parameters such as its home location, type of drones, available sensors, and sensor noise. Depending on the type of system, this can either be created manually, by implementing the respective glue code and setup scripts for the simulator, or, in the case of an existing system model (where an MDE approach is employed), can even be generated automatically using model-to-code-transformation.
 
Finally, when the stories are executed in the simulation environment, respective test data is collected and test traces are created. 
This in turn requires to establish runtime monitors, in accordance with the V\&V model. Depending on the properties to be validated during simulation, different information needs to be collected, both from the simulation environment and the SuT itself.

\subsection{Test Analysis \& Reporting}
Finally, when the scenarios are executed, the test traces need to be analyzed and a corresponding \emph{Test Report} needs to be generated. In the case of simulation-based testing, current testing analysis capabilities are quite limited.  Simulation tools only indicate whether a simulation has finished successfully or failed. They do not provide insights into (a) why a specific simulation scenario failed, or (b) whether all properties of the test were maintained during the simulation. For example, whether a sUAS maintained its maximum allowed deviation from the planned path, or if battery consumption stayed within the threshold limit. 
Similarly, for test scenarios executed in the field, typically, only a limited amount of information is available, to what extent tests have passed or failed. As a result, developers must rely on SuT logs and are responsible for both collecting and analyzing simulation data. As a result, the lack of specialized tools for post-simulation data collection and analysis presents a significant barrier.



\subsection{Treacebility \& Closing the Loop}
One of the main advantages of a structured and well-defined testing and simulation process lies in the fact that these tasks can largely be automated.
With runtime monitors and subsequent analysis in place, these issues can be automatically detected and reported.
Test results furthermore serve as inputs for any safety assurance process such as Safety Assurance Cases.
Typically, a SAC represents a set of claims, arguments, and finally, evidence created to justify that a system is sufficiently safe, satisfying its requirements~\cite{denney2015dynamic}. For example, successful simulation test results, such as a mission completed under 23mph wind conditions, can serve as evidence supporting the safety claim that the system can navigate safely in extreme windy conditions. In order to close the loop, simulation results and safety analysis artifacts further provide feedback to improve or modify the requirements model, V\&V model, or the test model.

%% file: tab_overview.tex
\begin{table}[t!]
\caption{Overview of artifacts and models part of our proposed process.}
\label{tab:overview}

\addtolength{\tabcolsep}{-0.6pt}
\begin{tabular}{@{} p{1.65cm}p{7.1cm}@{}}

 \toprule

\textbf{Artifact}       &  \multicolumn{1}{c}{\textbf{Contents/Description}} \\ 
\midrule

Req Model & Reqs. for the SuT. \emph{[formalized, structured, or NL]}\\

V\&V Model  & Properties of the SuT that need to hold and should be tested. \emph{[formalized, e.g., temporal logic]}\\ \midrule

Test  Model  & Scenario descriptions that must be tested, \lof requirements. \emph{[state machines, structured text]}\\ 

Test Exec. Req.  & Required capabilities of the execution environment (e.g., required simulator features). \emph{[structured text, e.g., XML/JSON]}\\ 

Fixture  & Implementation and setup for SuT to execute tests. \emph{[code, setup scripts, parameterization files]}\\ 

Test Story  & "Materialization" of a concrete test to be executed specific to the test context (e.g., simulator execution, or field test guide "handbook").\emph{[executable scripts and properties, guidelines]}\\ \midrule

Test Trace  & Collected runtime testing data. \emph{[SuT log files, died test protocol]}\\ 

Test Report  & Aggregated and processed test results. \emph{[structured and visualized results, automated test verdicts]}\\

 \bottomrule
\end{tabular}

\vspace{-0.8em}
\end{table}

%% file: sec_04_instance.tex
\section{Application Example}
\label{sec:instance}

In the previous section, we have outlined our proposed process, its main artifacts, and activities throughout the testing cycle. To assess the general feasibility of the process and its constituent steps, we conducted a preliminary evaluation where we applied it to testing activities of our own sUAS system which we have been developing for several years, as part of our research. As described, we have faced issues and shortcomings with existing testing and simulation tools, and in this preliminary application of the process, we investigate if and to what extent these can be alleviated with this more structured set of activities and artifacts. 

\subsection{Testing sUAS Systems} 
We selected two exemplary requirements related to sUAS executing a mission in extreme weather conditions and collision avoidance using the EARS syntax~\cite{mavin2009easy}.\\
\emph{R1: A sUAS shall complete a flight with multiple waypoints in wind gusts }\\
\emph{R2:A sUAS shall complete a flight  without colliding with stationary objects, the terrain, or other aircraft}\vspace{-0.01em}

As discussed in the previous section, the V\&V model should include test properties relevant to the SuT, as well as environmental conditions. Therefore, we focus on incorporating conditions, such as wind gusts of 23mph in the case of R1, and an obstacle density of 0.4, in the case of R2, specifying that 40\% of the environment is occupied by obstacles. Next, we specify the test model using a state transition diagram (represented in a textual notation~\cite{gaspary2001distributed}) as it effectively captures the dynamic behavior of the SuT under multiple conditions.

For both,  R1 and R2, we first considered testing at \lof-1 using AirSim \cite{shah2018airsim} as a  simulation environment offering robust support for simulating various weather conditions, such as wind, and rain, and providing essential tools for inserting static obstacles into the scene. These features of the simulator allow us to initially conduct testing at \lof-1 while maintaining test execution requirements before moving to HITL tests with the flight controller (\lof-2), and ultimately field tests (\lof-3). Specifically, the test execution requirements for R1 and R2 at \lof-1 include adding a digital model of the real world to the simulation environment, as well as configuring the wind's origin, direction, and velocity in the simulation environment. As a result, to configure the environmental aspect of our test story, we can specify (a) loading the Unreal Engine plugin for geospatial data~\cite{Cesiumfo86:online}, (b) wind characteristics through AirSim that allows simulating wind in the 3D environment. Next, to simulate sUAS as per our test model in a 3D environment, we first need to establish a connection between the SuT and the simulation environment by specifying a connection string in AirSim. Second, we specify the sUAS mission as a JSON message. sUAS SuT commonly employ standard JSON messages to specify a mission, listing waypoints and take-off and landing locations \cite{al2023configuring}.

In addition, based on the V\&V specification, we can develop custom runtime monitoring scripts to trace the simulation run and collect relevant data. As a result, the collection of environment, SuT connection, sUAS mission, and runtime monitoring specifications form our test story. Since the entire test story is available as a formal specification,  the test story can be executed automatically multiple times, depending on the developers' needs. Finally, the generated trace contents and their automated analysis artifacts are linked to the requirement and ultimately to safety artifacts, such as an SAC, that argue whether the system is safe under specific environmental contexts. We provide exemplary snippets of artifacts from our sUAS system in  \citetable{example}.
\input{tab_examples}

After the successful execution and validation of a test story in the simulation environment (\lof-1), developers can plan to execute the same test story in higher \lof. For instance, rather than using the SITL flight controller (PX4) of the sUAS, we can execute the PX4 process on a Pixhawk board that is directly connected to the simulation environment. This HITL setup will generate actuator signals directly from the real hardware, bringing the fidelity of our simulation a step closer to reality. Since AirSim supports HITL simulations, we can reuse our test story specs and custom scripts to test the requirements at \lof-2 seamlessly. 

Following the successful execution of a test story at \lof-2, developers can finally proceed to execute the same test story in the field (\lof-3). During field tests, developers need to ensure that the testing site complies with the environmental requirements such as wind velocity in the area and obstacle density, and accordingly plan the SuT sUAS flight path. Developers should be able to reuse the same monitoring scripts used in \lof-2 to verify the test properties. Furthermore, developers should also collect hardware logs, such as temperature readings, power supply voltage, GPS strength, and radio signal reception, to provide additional context to the test artifacts. This process of gradually moving from a lower to a higher fidelity testing environment enables the comparison of test results between levels and provides feedback to the lower fidelity testing environment for refinements.




\subsection{Discussion}

Our preliminary experimental application has demonstrated that our process is adaptable to various testing activities, and suitable for both simulation and field testing. Based on these initial insights, a structured test model with clear test execution requirements provides a formal framework to identify tests, execution context, and monitoring properties. Further, a detailed test "handbook" derived from the model, outlining steps and actions, results in documented, reproducible tests.

\noindent{\textbf{Generalizability}}: Concerning the application of our process beyond sUAS systems, further evaluations need to be conducted. However, based on our experience with robotic applications, we are confident that the process is also applicable to systems with similar characteristics. 
For example, testing an autonomous robot in a smart warehouse with requirements such as  \emph{"When a human worker enters the path of the robot, the Computer Vision (CV) component shall detect the human and avoid a collision"} could follow a similar testing process. The V\&V model in this instance contains properties related to the detection speed, accuracy, and avoidance states.  The simulation environment must include a 3D model of a worker that reflects diverse human characteristics -- age, sex, height, skin tone, and clothing preferences -- and mimics typical warehouse worker behaviors to meet execution requirements. Such execution requirements, for example, result in Gazebo~\cite{gazebo} being selected as simulation environment for certain test stories in conjunction with CV tests. Using a ROS-based application, used for robotic applications, a monitoring system can easily be established to collect and log runtime data and analyze the results of the simulations as evidence for an SAC about safe navigation in the presence of a human worker.

\noindent\textbf{Future Evaluation Plans:} We are actively pursuing two directions. First, we plan on applying an extended and iteratively refined process (cf. research roadmap) to a specific application scenario from our sUAS system (e.g., a search-and-rescue or surveillance scenario) where we fully track the testing process end to end, and subsequently apply it to a second system in the robotics domain, as sketched out above. Secondly, we plan to conduct a user study with sUAS developers to evaluate this process through both simulation and field testing, aiming to assess the effectiveness of various modeling formalisms.









%% file: tab_examples.tex
\begin{table*}[]
\centering
\caption{Example Requirements and model contents.}
\label{tab:example}
\scriptsize
\addtolength{\tabcolsep}{-0.5em}
\def\arraystretch{1.6}
\begin{tabular}
{L{0.35cm}|L{2.3cm}|L{3.3cm}|L{2.1cm}|L{3cm}|L{2.5cm}|L{3.2cm}}
\toprule
\textbf{Req.} & \textbf{V\&V Model}  & \textbf{Test Model}   & \textbf{Exec. Req.}       & \textbf{Test Story}   & \textbf{Trace Contents}    &  \textbf{Safety Claim}  \\ 

\midrule
R1 & 
\texttt{wind\_gusts \textless{}= 23mph \& deviation \textless 5\% }&
 \multirow{2}*{

\begin{minipage}[t]{3.2cm}
\textsl{FinalState:} \texttt{mission\_finished}\newline
\vspace{0.2em}
\textsl{State}\texttt{ active ``prearm-checks successful''}\newline
\textsl{GoToState} \texttt{ready-for-takeoff}\newline
\vspace{0.4em}
\textsl{State}\texttt{ ready-for-takeoff ``mission-assigned''} \newline \textsl{GoToState} \texttt{request-takeoff}\newline 
\texttt{...}
\end{minipage}
}
& 
$\bullet$ Geospatial Data\newline
$\bullet$  Wind model \newline(Vel, Dir, Coord)\newline &
$\bullet$ Simulator: AirSim \newline
$\bullet$ Sim Env: Wind Config \newline
$\bullet$ SuT sUAS Mission: JSON\newline
$\bullet$ SuT-Sim Conn.: String\newline
$\bullet$ Runtime Monitors: Scripts
\newline

& $\bullet$ Flight Path\newline
$\bullet$ sUAS IMU States \newline
$\bullet$ SuT States \newline
$\bullet$ Resultant Forces on sUAS \newline
&

 \multirow{2}*{

\begin{minipage}[t]{3cm}
\emph{Claim:} sUAS does not collide with objects in dense environment when flying under 23mph gusty winds \newline

\bigskip

$\bullet$ \emph{SC1:} sUAS remains stable under gusty winds of up to 23 mph

$\bullet$ \emph{SC2:} sUAS avoids collision in dense environment (buildings, ground, other sUAS)\newline
\end{minipage}
}

\\   \cline{1-2}\cline{4-6}

R2&
\texttt{
obs\_density=0.4 \& col.count =0 \& miss.success = True} &

& 
$\bullet$ Geospatial data \newline
$\bullet$  Wind model \newline(Vel, Dir, Coord) \newline
$\bullet$ Obstacle \newline (Type, Location)
&
$\bullet$ Simulator: AirSim \newline
$\bullet$ Sim Env: Cesium (Unreal) \newline
$\bullet$ SuT sUAS Mission: JSON\newline
$\bullet$ SuT-Sim Conn.: String\newline
$\bullet$ Runtime Monitors: Scripts
& $\bullet$ Landing Location\newline
$\bullet$ sUAS IMU States \newline
$\bullet$ SuT States \newline
$\bullet$ Proximity to Obstacles \newline
&  \\

\bottomrule
\end{tabular}
\vspace{-2em}
\end{table*}

%% file: sec_05_roadmap.tex
\section{Research Roadmap}
\label{sec:roadmap}
As part of our ongoing work in this area, we have identified six primary research directions, related to both foundational questions (cf.~traceability, test representation, and model management and consistency) and technical challenges (cf.~simulator integration, mapping simulator capabilities, and quantifying the simulation-reality gap) as elaborated in the following.

$\bullet$ \textbf{Traceability:}
Traceability emerges as a significant factor, especially in the context of safety-critical applications and safety analysis. Traceability is essential for conducting thorough risk assessments to detect potential hazards and their causes~\cite{agrawal2023leveraging ,hawkins2013assurance,denney2015dynamic}. Nonetheless, it is imperative to provide diverse credible evidence that hazards can be effectively mitigated. However, to accomplish this, the procedure must be supplemented by an appropriate tracing strategy and traceability information model (TIM) that accounts for these connections. This is also emphasized by Mäder~\etal~\cite{mader2013strategic}, particularly focusing on the importance of reliable traceability, which they argue is lacking in practice. We aim at leveraging inter- and intramodel linking for tracing requirements~\cite{galvao2007survey} to their realization in the SuT and the tests targeting these.

$\bullet$ \textbf{Model Homogeneity and Consistency:}
Aside from traceability, linking the different models is crucial, to ensure consistency across requirements, the V\&V, and the test model. This stems from the circumstance that different modeling formalisms are employed at the various stages, resulting in heterogeneous models, or even code snippets (e.g., when validating simulation results). Yet, the different models share identical properties and require trace links among them (cf.~Traceability). If not properly managed, inconsistencies can be easily introduced throughout the whole process~\cite{ractiu2023taming}. In traditional model-based development (MBD), commonly one modeling approach is used, however, the diverse nature of process levels and testing context (e.g., formal functional requirements testing vs.~automated simulator testing vs.~manual field testing) calls for support of a diverse set of models. As in the case of traceability, we leverage techniques from MBD, specifically ultra-larger scale heterogeneous model synchronization~\cite{hill2008towards}.

$\bullet$ \textbf{Representation of Test Stories:}
A structured and flexible description of different test stories is required. While model-based, and scenario-based testing are well-established techniques, solutions are typically limited to a narrow scope and usually target one execution environment; in our case, we have at least two - simulation and reality. Consequently, test stories need to cover a wide range from formal tests (cf.~checking properties during simulation execution), to manually performing test steps at a field test with human interaction. For this purpose, we proposed the investigation of domain-specific languages (DSLs) tailored towards test story specification, similar to the classic notion of a movie script which describes the environmental setting (cf.~simulator or reality), the activities to happen (cf.~instrumentation of the SuT), and eventually a test oracle for subsequent analysis. Capitalizing on DSLs substantially facilitates automation, allowing for intelligent test selection by covering a wide range of conditions with potentially limited resources~\cite{zech2012generic}.

$\bullet$ \textbf{Extended \lof:}
The Level of Fidelity description provides the ability to incorporate different testing contexts, with different levels of detail and realism during test execution. 
Similar concepts are, for example, applied in the construction domain where Building Information Modelling (BIM) models~\cite{succar2009building} comprise multiple levels of detail (LoD), referring to the fidelity of the model w.r.t.~the information provided therein. For example, BIM LoD clearly specify the requirements of what a model needs to contain at 
LoD 100, where only 2D symbols and basic elements exist, up to LoD 500 where the actual functions of elements in real buildings are described. 

In our context, the \lof provides an iterative refinement of test specification and execution, where each level adds an additional degree of assurance. Currently, the four proposed levels (\lof[0] to \lof[3], cf.~Sec.~\ref{sec:solution}) are, however, still rather coarse-grained. One dedicated area for 
further research thus is a comprehensive description and specification of the proposed levels. For example, Software-In-the-Loop (SITL) tests may be further refined into multiple \lof based on simulator capabilities. Simulators with realistic environmental models or physics engines would rank higher within \lof-1 than those with minimal environmental or low-fidelity physics simulations. 

%




$\bullet$ \textbf{Mapping Simulator Capabilities:}
Simulation environments currently cover a wide range of features and functionalities with various \lofs. While in the automotive domain, highly sophisticated, feature-rich simulation 
environments exist (e.g.~CARLA), other areas still lack sophisticated simulation support. Particularly for sUAS, there is a need for standardized simulation support that enables 
methodological testing and validation~\cite{agrawal2023requirements}. To bridge the gap between simulation and testing, we are currently conducting an extensive literature analysis on sUAS simulation environments, with the goal of documenting commonalities and variability alongside the different features that are available, similar to the automotive 
domain~\cite{thorn2018framework}. This will not only deliver a better overview of existing solutions but --- more importantly --- provide the foundation for (automatically) mapping 
different \lofs.
We envision a structured description of capabilities for simulation environments, similar to a feature model. For example, AirSim, based on Unreal Game Engine Technology, provides features for modeling and rendering 3D 
environments, including geospatial and weather data. This information can in turn be used to inform the \lof of a certain test story (or determine the desired \lof of a test story to be 
executed in a specific execution context), and automatically generate the respective setup and configuration scripts for test execution. As a result, simulation runs can be executed in a CI/CD DevOps process, allowing traceability between implemented requirements, simulator configurations, and test results.

$\bullet$ \textbf{Simulator-to-Reality Gap:}
Finally, the simulator-to-reality gap has long been recognized as a critical issue in robotics research~\cite{jakobi1995noise}. It describes the mismatch of simulated and real physical-law performances caused by the inaccurate representation of the real environment in simulation~\cite{salvato2021crossing}. In our case, this issue becomes evidently pertinent in moving testing from \lof[1] to \lof[3]. Our approach however is not designed to reduce the simulator-to-reality gap but rather, as of its theoretical 
foundation, provide a tool to quantify this gap by correlating simulated and real execution traces. Put differently, our framework provides the measurement basis for aligning the fidelity of 
a simulation with the fidelity of reality, put simply, in how far results from simulation map to reality without loss of granularity.

\bigskip

The above-mentioned six directions provide ample opportunities for improving the current state of the art in simulation-based validation of sUAS. While some research areas, like traceability, are well established, there is a need to explore automated approaches for supporting the traceability of simulation artifacts to assess and analyze sUAS safety. Additionally, integrating concepts from other domains, such as the proposed LoF simulator classification, requires an in-depth analysis of current simulator capabilities and a mapping to various types of domain- and system-specific requirements. Furthermore, our proposal is model-driven, allowing us to maintain consistency throughout the entire testing process, transferable across multiple domains, and providing opportunities to automate the process to the maximum extent possible.







%% file: sec_06_conclusion.tex
\section{Conclusion and Future Work}
\label{sec:conclusion}
Software Testing is a well-established research area, with mature processes, tools, and techniques.
However, in the sUAS and CPS domain, despite the existence of simulation environments commonly employed for testing, there is still a lack of support for executing a well-defined testing scenario, with an overall process for combining different types of tests, environmental factors, and properties. This often results in situations where systems are tested in inadequate simulation environments that do not adequately capture the real-world complexity, are only tested in a limited range of scenarios, and lack standard safety analysis processes. In this paper, we presented an initial framework for validating CPS, specifically focusing on sUAS and robotic applications, demonstrating its suitability by applying it to two different types of CPS. Based on our initial observations and insights, we have compiled a research roadmap with five ongoing research activities related to model management and consistency, simulator capabilities, traceability, and structured test scenario representation.